\documentclass[iop]{emulateapj}
\usepackage{epsfig}

\def\mnras{Mon.\ Not.\ R. Astron.\ Soc.}
\def\apj{Astrophys.\ J.\ }
\def\aj{Astron.\ J.}

\def\nat{Nature}

\def\aap{Astron.\&Astrophys}

\def\mcut{m_{\mathrm{min}}}

\def\msun{M_\odot}

\newcommand{\tcad}{t_{\mathrm{cad}}}

\newcommand{\mlim}{m_{\mathrm{lim}}}
\newcommand{\Afs}{A_{\mathrm{fs}}}
\newcommand{\GAIA}{{\it Gaia}}
\newcommand{\LSST}{{\it LSST}}
\newcommand{\WFIRST}{{\it WFIRST}}
\newcommand{\Nomad}{{nomad}}
\newcommand{\Nomads}{{nomads}}

\newcommand{\alphapl}{\alpha_{\mathrm{nm}}}
\newcommand{\alphabd}{\alpha_{\mathrm{bd}}}
\newcommand{\Nnm}{N_{\mathrm{nm}}}
\newcommand{\NMS}{N_{\mathrm{MS}}}

\usepackage[usenames]{color}

\shorttitle{Nomads of the Galaxy}
\shortauthors{Strigari et al.}

\begin{document}

\title{Nomads of the Galaxy}


\author{Louis E. Strigari\altaffilmark{1}, 
        Matteo Barnab\`e\altaffilmark{1}, 
        Philip J. Marshall\altaffilmark{2}, 
        Roger D. Blandford\altaffilmark{1}}

\altaffiltext{1}{Kavli Institute for Particle Astrophysics and
  Cosmology, Stanford University, Stanford, CA 94305 USA}
\altaffiltext{2}{Department of Physics, University of Oxford, Keble
  Road, Oxford, OX1 3RH, UK}

\begin{abstract}
We estimate that there may be up to $\sim 10^5$ compact 
objects in the mass range $10^{-8} -10^{-2} \msun$
per main sequence star that are unbound to a host star
in the Galaxy. We refer to these objects as {\it nomads}; 
in the literature a subset of these are sometimes called free-floating or 
rogue planets. Our estimate for the number of Galactic nomads is consistent with a 
smooth extrapolation of the mass function of unbound objects above the Jupiter-mass scale, 
the stellar mass density limit, and the metallicity of the interstellar medium. We analyze the 
prospects for detecting nomads via Galactic microlensing. 
The Wide-Field Infrared Survey Telescope ({\WFIRST}) will measure the number of nomads 
per main sequence star greater than the mass of Jupiter 
to $\sim 13\%$, and the corresponding number greater than the mass of Mars 
to $\sim 25\%$. All-sky surveys such as {\GAIA} and {\LSST} 
can identify nomads greater than about the mass of Jupiter. We suggest 
a dedicated drift scanning telescope that covers approximately 100 square degrees 
in the Southern hemisphere could identify nomads as small as $10^{-8} \msun$ via microlensing
of bright stars with characteristic lightcurve timescales of a few seconds. 
\end{abstract}

\keywords{gravitational lensing -- planets and satellites: general -- Galaxy: general}

\maketitle

\section{Introduction} 

The recent years have witnessed a rapid rise in the number of known planetary mass
objects, $\lesssim 0.01 \msun$, in the Galaxy. 
Searches for exoplanets from radial velocities
find that $\sim 30-50\%$ of GK dwarf stars
have planets greater than the mass 
of Neptune within periods $< 50$ days~\citep{2011arXiv1108.5842W}. 
Transiting searches find that $\sim 15\%$ of main sequence dwarfs 
are orbited by short-period planets at
less than four Earth-radii
~\citep{2010Sci...327..977B}. Direct imaging and microlensing 
have also now started to uncover planets bound to host stars
~\citep{2010arXiv1002.0332G}. 

Much less is known about the population of $\lesssim 0.01 \msun$ objects that are 
not bound to a host star. Several candidate unbound objects have been imaged 
in star clusters with mass possibly as small as a few times that of Jupiter, $10^{-3} \msun$
(e.g.~\cite{2006Sci...313.1279J,2007A&A...470..903C,2009A&A...506.1169B}). However, the origin
of these objects is uncertain; they may have formed directly in the collapse of the molecular
cloud~\citep{1976MNRAS.176..483R},
or have been ejected from their birthplace around a host star via a dynamical interaction
~\citep{2000ApJ...536L.101B}. 
Free-floating objects at the Jupiter mass and below have been difficult to find via these
methods. 

Microlensing, however, does provide
a way---and perhaps the only way---to detect 
objects below the deuterium-burning mass limit that are not bound to a host star. 
In a recent survey of the Galactic bulge, the MOA-II
collaboration~\citep{Sumi:2011kj} reported the discovery of
planetary-mass objects either very distant from their host star ($\sim
100$ AU) or unbound from a host star entirely.  This detection was
obtained from analysis of the timescale distribution of the
microlensing events, which showed a statistically-significant excess
of events with timescale $\lesssim 2$ days as compared to a standard
Galactic model with a stellar mass function cut-off at the low mass
end of the brown dwarf regime.  The mass function of this new
population of objects 
can be described (for illustration) by a $\delta$-function with a
best fit near the Jupiter mass. These
results tell the surprising story that objects greater than about the mass
of Jupiter are approximately twice as numerous as both main sequence stars 
and planets bound to host stars.

Though their existence is established, the origin of these unbound objects
is far from clear. Do they form a continuation of the low end brown dwarf mass 
function near the deuterium burning mass limit, or did they form as a distinct 
population of objects ejected from their original host stars? Because of their
uncertain origin and their present status, we prefer to  
refer to objects with mass $< 0.01 \msun$ that are not bound to a host star 
as {\it Nomads}; in the literature they have been also referred to
as rogue or free-floating planets. The name ``{\Nomad}" is invoked to include 
that allusion that there may be an accompanying ``flock," either in the form of 
a system of moons~\citep{2007ApJ...668L.167D} or in its own 
ecosystem. Though an interstellar object might seem an especially inhospitable 
habitat, if one allows for internal radioactive or tectonic heating and the 
development of a thick atmosphere effective at trapping infrared heat
~\citep{1999Natur.400...32S,2011ApJ...735L..27A}, and recognizes that 
most life on Earth is bacterial and 
highly adaptive, then the idea that interstellar (and, given the prevalence of 
debris from major galaxy mergers, intergalactic) space is a vast ecosystem, 
exchanging mass through chips from rare direct collisions, is intriguing with 
obvious implications for the instigation of life on earth.

Understanding the bounds on the {\Nomad}
population, and the prospects for detecting them with microlensing surveys, 
is the focus of this paper. In particular, what is the number
and mass density of {\Nomads} in the Galaxy? 
What are the bounds on the minimum mass of a {\Nomad}, and what is the lightest 
detectable {\Nomad}?  How well can we measure the {\Nomad} 
mass function, and how does this compare to the low mass brown dwarf mass 
function? And can we independently measure the mass function of {\Nomads}
in the bulge and in the disk?

We will show that a dedicated space-based survey of the
inner Galaxy, such as the proposed Wide-Field Infrared
Survey Telescope ({\WFIRST}\,\footnote{http://wfirst.gsfc.nasa.gov/}), 
will measure the number of nomads 
per main sequence star greater than the mass of Jupiter 
to $\sim 13\%$, and the corresponding number greater than the mass of Mars 
to $\sim 25\%$. 
We also show that large scale Galaxy surveys, in particular the 
{\GAIA}\,\footnote{http://sci.esa.int/science-e/www/area/index.cfm?fareaid=26}
mission and the Large Synoptic Survey Telescope 
({\LSST}\,\footnote{http://www.lsst.org/lsst/}), will be
sensitive to {\Nomads} greater than about the mass of Jupiter without changing
their proposed observing plan. Further, the back-to-back 15 second exposures
in the planned design of  {\LSST} will allow for limits at least to be placed on nomads
near the mass of Pluto, $\lesssim 10^{-8} \msun$. As an extension, we suggest that a 
dedicated drift scanning telescope
could identify {\Nomads} as small as $10^{-8} \msun$ via microlensing
of bright stars with characteristic lightcurve timescales of a few seconds. 

This paper is organized as follows. In Section~\ref{sec:population} we
estimate the number of {\Nomads} in the Galaxy. In 
Section~\ref{sec:event_rate} we calculate the event rate of {\Nomads}
in microlensing surveys. In Section~\ref{sec:projections} and 
Section~\ref{sec:efficiency} we outline methods for measuring the {\Nomad} population
and simulating detection efficiencies.  In Section~\ref{sec:surveys}
we present the results of these projections. In Section~\ref{sec:drift} we
postulate a survey for short timescale {\Nomads} via a drift scanning
telescope. In Section~\ref{sec:conclusion} we summarize our conclusions.

\section{The Nomadic population}
\label{sec:population} 

We begin by setting up the model for the {\Nomad} population.  Objects 
with mass $< 10^{-2} \msun$ are believed to originate from two distinct processes. Between 
the Jupiter mass and the deuterium-burning mass, many of these objects 
may form similar to stars by gravitational fragmentation. Below Jupiter masses, they likely 
are born in protoplanetary disks and dynamically-ejected during the 
evolution of the system. It is unknown from a theoretical perspective whether there
is a smooth continuation of the mass function at the dividing
mass that separates these populations. 

In light of these uncertainties, we choose a simple broken power-law model for the 
{\Nomad} mass function, $dN/dM \propto M^{-\alpha}$, which is a smooth continuation of
the mass function at higher masses, 
\[
\alpha \equiv \left \{ \begin{array}{lcl@{\,}c@{\,}l} 
  \alphapl   & {\rm for} \, &
         10^{-8} & M/\msun & < 0.01 \\ 
  \alphabd   & {\rm for} \, & 
         0.01 \le & M/\msun & < 0.08 \\
  \alpha_2 & {\rm for}  \, &
        0.08 \le & M/\msun & < 0.70 \\
 \alpha_1 & {\rm for} \, &
         0.70 \le & M/\msun & .
\end{array} \right. \\
\]
In addition to these power law slopes, we define the minimum cut-off in the {\Nomad} mass function
as $\mcut$, the number of objects in the {\Nomad} mass regime as
$\Nnm$, and the number of main sequence stars in the mass
regime of $0.08-1 \msun$ as $\NMS$. From the latter two
quantities we define the ratio $\beta \equiv \Nnm/\NMS$. 

For the above parameterization, empirical bounds over the entire nomad range may be motivated
from several considerations. First, the mass function of the lowest mass nomads that we consider 
may be estimated from bounds on the population of Kuiper Belt objects
(KBOs). We start from the distribution of diameters of KBOs determined
in~\citet{Bernstein2004}.  At the high diameter end, $D \gtrsim 100$
km, the KBO distribution scales as $dN/dD \propto D^{-4}$. Below the
break radius of $D \sim 100$ km, where collisional effects are
believed to be important, the KBO distribution flattens, $dN/dD
\propto$ const. Assuming that the bodies of the outer Solar System
have approximately constant mass density of $\sim 1$ g cm$^{-3}$, the mass function scales as
$dN/dM \propto m^{-2}$ above $10^{-12} \msun$, while below the mass
function scales as $dN/dM \propto m^{-2/3}$. 

At the highest mass end, corresponding to approximately several times the mass of 
Jupiter, there is evidence that nomadic objects in open clusters 
constitute a smooth continuation of the brown dwarf mass function at higher
masses with $\alpha = 0.6$~\citep{2007A&A...470..903C}.
Further there may be a turnover in the 
mass function below $\sim 6$ times Jupiter mass~\citep{2009A&A...506.1169B}, 
though these results are subject to systematic uncertainties on 
the masses of the objects and the small number of objects known. 

In comparison to these results from direct imaging, microlensing observations 
more strongly constrain the nomadic mass function,
in particular at the high mass end. The microlensing results from 
~\citet{Sumi:2011kj} find that the equivalent best-fitting slopes and 
one-sigma uncertainties are $\alphapl = 1.3_{-0.4}^{+0.3}$ 
and $\alphabd = 0.48_{-0.27}^{+0.24}$.
In~\citet{Sumi:2011kj} the minimum
mass was taken to be $\mcut = 10^{-5} \msun$, though given the
cadence of the survey they are insensitive to values of $\mcut$ at
this mass scale and below. Taking $\mcut = 10^{-5} \msun$ and these
best-fitting slopes implies $\beta \simeq 5$. Extrapolation
down to below Earth mass scales, $\mcut = 10^{-6} \msun$, yields
$\beta \simeq 10$, and further extrapolation down to $\mcut = 10^{-8} \msun$
yields $\beta \sim 60$. Intriguingly for a continuous power law extrapolation 
down to $\sim 10^{-15} \msun$, the number of nomads per star approaches the
bound on the abundance of interstellar comets
~\citep{2005ApJ...635.1348F,2011AJ....141..155J}, 
and the corresponding {\Nomad} mass density is $\sim 1\%$ of the oxygen mass 
density in the interstellar medium~\citep{2006ApJ...639..929B}. 

Assuming the above parameterization of the mass function, $\alphapl$
is negatively correlated with $\alphabd$ from microlensing
observations.  For example, for $\alphabd = 1$, the 95\% c.l. lower
limit on the {\Nomad} slope is $\alphapl = 0.5$. For this combination of
slopes, extrapolating down to $\mcut = 10^{-8} \msun$ implies that
$\beta \gtrsim 1$.  On the other hand for $\alphabd = 0$,
the~\citet{Sumi:2011kj} 95\% c.l. upper limit on the slope 
is $\alphapl = 2$. In this case assuming $\mcut
= 10^{-5} \msun$ implies $\beta \sim 50$. Extrapolation down to $\mcut
= 10^{-6} \msun$ implies an order of magnitude increase in $\beta
\simeq 700$, while extrapolation down to $\mcut = 10^{-8} \msun$ 
implies $\beta \simeq 10^5$. 

The above estimates indicate that, when fixing to the measured
abundance of {\Nomads} at $\gtrsim 10^{-3} \msun$ and smoothly
extrapolating to lower masses, there is several orders of magnitude
uncertainty on the {\Nomad} abundance.  
For an appropriately large ratio of the number of {\Nomads} to main
sequence stars, the {\Nomad} mass function is constrained by the limits on
the number of compact objects in the Galactic disk and halo. For
masses $\gtrsim 10^{-7} \msun$, the abundance of compact objects in
the halo is constrained by upper limits on MACHO dark
matter~\citep{Tisserand2007}.  These bounds indicate that compact
objects of mass $\gtrsim 10^{-7} \msun$ comprise $\lesssim 10\%$ of
the Galactic dark matter halo.  For more local measurements, 
a less
stringent limit arises from the stellar mass density of the Galactic
disk, which we take to be $\rho_0 = 0.1 \msun$
pc$^{-3}$~\citep{Holmberg:1998xu}.  
As an example, if we assume that
the entire local mass distribution of the disk is comprised of objects
at the mass scale $10^{-8} \msun$, the bound $\rho_0 = 0.1 \msun$
pc$^{-3}$ corresponds to an upper limit of $10^6$ compact objects per
main sequence star.  Masses of compact objects at these scales and
below may be probed by future short cadence microlensing observations
with the Kepler satellite 
\citep[see][and the discussion below]{Griest:2011av}.

\begin{figure}
\begin{center}
\begin{tabular}{c}
{\includegraphics[height=7.0cm]{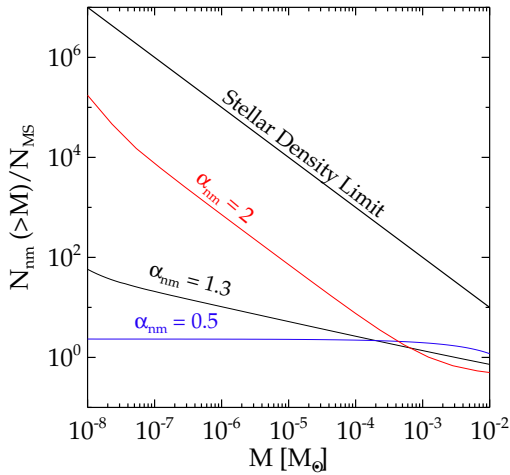}} \\
\end{tabular}
\end{center}
\caption{Number of nomads greater than a given mass scale, $N_{nm} (>M)$,
  relative to the number of main sequence stars,
  $\NMS$. Three different slopes for the {\Nomad} mass function
  are labeled, $\alphapl = 2,1.3,0.5$. The upper black curve is
  determined assuming that objects at that mass scale have a density
  of $\rho_0 = 0.1 \msun$ pc$^{-3}$.
\label{fig:mf}
}
\end{figure}

Predictions for the number of {\Nomads}, in comparison to constraints on
the local mass density, are summarized in Figure~\ref{fig:mf}.  This
shows the number of nomads greater than a given mass scale, $N(>M)$,
relative to the number of main sequence stars,
$\NMS$. Four different slopes for the {\Nomad} mass function
are labeled, $\alphapl = 2,1.3,0.5$, which have corresponding values
for the slope of the brown dwarf mass function of $\alphabd =
0,0.5,1$. The upper limit, indicated as the diagonal line, is
determined assuming that objects at that mass scale have a density of
$\rho_0 = 0.1 \msun$ pc$^{-3}$. For $\alphapl \ge 2$, the total mass of the nomad population 
is dominated by the lowest mass objects. 
 
\section{Event Rates} 
\label{sec:event_rate}

In this section we calculate the microlensing event rates
from the {\Nomad} population.  We begin by establishing the definitions and
the Galactic model parameters, and then use this model to predict the
timescale distribution of events and the integrated number of events
detectable. 

\subsection{Definitions} 

We employ standard microlensing formalism. The distance to the source
star is $D_S$, the distance to the lens is $D_L$, and the mass of the
lens is $M$.  The Einstein radius is $R_E^2 = (4G/c^2) M D_L
(D_S-D_L)/D_S$, and the Einstein crossing timescale is $t_E = R_E
/v$. The amplification of a source star is $A(u) =
(u^2+2)/(u\sqrt{u^2+4})$, where $u$ is the projected separation of the
lens and source in units of the Einstein radius.

To calculate microlensing event rates we use Galactic model~1
of~\citet{Rahal:2009yt}. This is characterized by an exponential thin
disk with a scale length of $R_d = 3.5$ kpc and scaleheight of $z_h =
0.325$ kpc,
\begin{equation}
\rho_d(R) = \rho_0 \exp \left[-(R-R_0)/R_d - |z|/z_h \right], 
\label{eq:rho_disk}
\end{equation}
where $R_0 = 8.5$ kpc, z$_h$ = 0.35 kpc.  We use the
bulge density distribution from~\citet{Dwek:1995xu}.  The lens-source
transverse velocity distribution, $f(v_l,v_b)$ is modeled as
in~\cite{Han:1995nt}, with $v = \sqrt{ v_l^2 + v_b^2 }$, where we
indicate as $v_l$ and $v_b$, respectively, the velocity along the
galactic longitude and latitude coordinates.

We determine the total event rate distribution by breaking the
lenses-sources into the disk-bulge, bulge-bulge, and disk-disk
components. We consider two different targets of source stars. First,
bulge stars in the direction of Baade's window, $(b,l) = (-3.9^\circ,
1^\circ)$, and second, sources distributed over all-sky. For the bulge
observations we can compare to observational determinations of the
optical depth \citep{Sumi:2002wg, Popowski:2004uv, Sumi:2005da,
  Hamadache:2006fw} and to the theoretical optical depth
calculations~\citep{Han:2003ws, Wood:2005au} by smoothly extrapolating
the rates from the {\Nomad} mass regime to the mass regime of main sequence
stars and remnants.  

\subsection{Finite source effects} 

Since we extrapolate the {\Nomad} mass function down to low mass 
scale, it is important to properly account for finite source effects
in the microlensing events. More specifically, we need to estimate by
how much the peak amplification of an event is reduced when $R_E$ is
of order the projected radius of the source star.  To estimate finite
source effects we take the source stars to have uniform surface
brightness, and for a given projected lens-source separation we
estimate the amplification as
\begin{equation} 
\Afs(u) = 
  \int_0^{2\pi} \int_0^{\rho_\star} d\phi \, \rho  d\rho 
  A( \sqrt{u^2+\rho^2-2u\rho \cos \phi} \, ),
\label{eq:finite_source}
\end{equation} 
where $\rho_\star = R_\odot/R_E$. For typical lens and source
distances, $D_L \simeq 5$ kpc and $D_S \simeq 8$ kpc, and assuming a
lens mass $M = 10^{-8} \msun$, the peak amplification is
$\Afs \simeq 1.1$. Though this is less than the standard
point lens-point mass amplification by $\sim 15\%$, surveys that we
consider below will still be sensitive to brightness fluctuations of
this magnitude.  Extrapolating further down to $M = 10^{-9} \msun$,
the peak amplification is only $\Afs \simeq 1.01$.  Though
extraction of events at this brightness may still be detectable, to
provide conservative estimates we restrict our analysis to lens masses
$\ge 10^{-8} \msun$.

\subsection{Bulge event rate}

The microlensing event rate per source star in a direction $(b,l)$ is
given by an integral over the lens-source transverse velocity
distribution, the lens density distributions~$\rho_L$ and the mass
function \citep{Griest:1990vu, KIraga1994},
\begin{eqnarray} 
\frac{d\Gamma(b,l)}{dt_E}
     & = & u_T \int_0^{D_S}  dD_L \int dv_l dv_b v  f(v_b,v_l) 
           \delta(t_E - R_E/v) \nonumber \\
     & \times & \int_{\mcut}^{\infty}  dM \zeta (M)  R_E \rho_L(l,b,D_L) . 
\label{eq:microlensing_rate}
\end{eqnarray} 
The mass function $\zeta (M)$ is normalized to the mean mass of the
lens population.  The optical depth is $\tau \simeq \pi \Gamma/2$,
where $\Gamma$ is the integral of the event rate distribution over all
$t_E$.  In Eq.~\ref{eq:microlensing_rate}, $u_T =
(A^2/\sqrt{A^2-1}-1)^{1/2}$, with $A = 1.34$ corresponding to $u_T=1$.
This corresponds to the event rate for source stars within a circular
area of one Einstein radius of the lens star.  This is a conservative
criteria that is appropriate for our analysis; event rates are
increased for $u_T >1$ when allowing for $A<1.34$.

\begin{figure*}
\begin{center}
\begin{tabular}{c}
{\includegraphics[height=6.5cm]{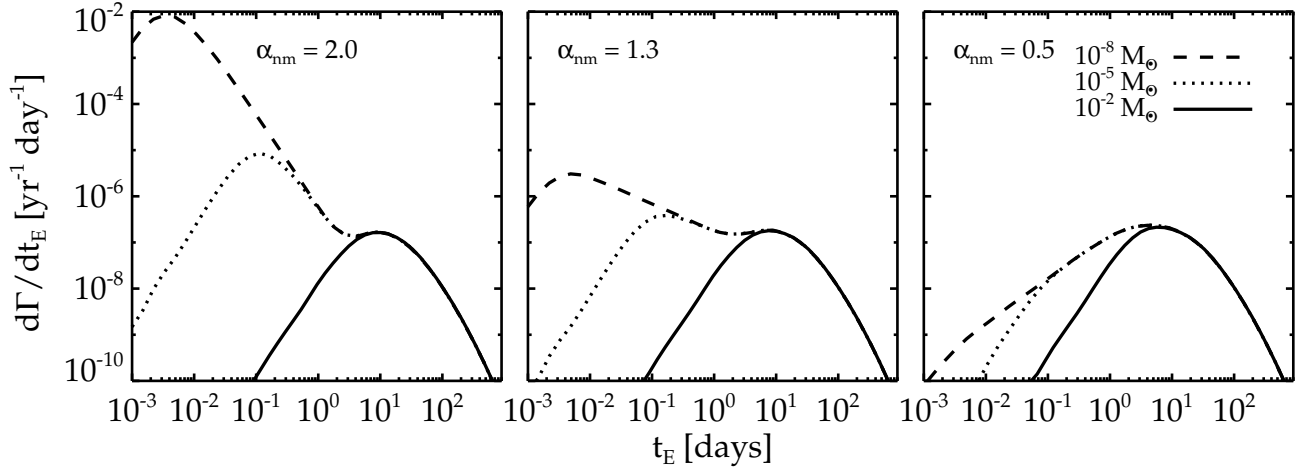}} \\
\end{tabular}
\end{center}
\caption{The event rate timescale distribution for several mass
  functions and cut-off masses. In all panels the solid, dotted, and
  dashed curves assume $\mcut = 10^{-2}, 10^{-5}, 10^{-8} \msun$,
  respectively.  The slopes of the mass function in the planetary mass
  regime ($\alphapl$) are indicated in each of the panels.
\label{fig:te}
}
\end{figure*}

Figure~\ref{fig:te} shows the event rate distributions in the
direction of the bulge, with each panel corresponding to a different
assumption for the slope of the {\Nomad} mass function. 
In all 
panels we take $\alpha_1 = 2.0$ and $\alpha_2 = 1.3$ for the main sequence
stellar mass function. Within each of the
panels there are three different assumptions for $\mcut$. The three
curves in all of the panels represent the sum of the event rate
distribution from bulge and disk lenses. For all curves in the middle
and right panels, the mean timescale of a microlensing event is
$\langle t_E \rangle \simeq 50$ days. However, for the curves in the
left panel, $\langle t_E \rangle$ depends strongly on $\mcut$ because
of the steep power law to low masses. Specifically for $\mcut =
(10^{-2}, 10^{-5}, 10^{-8}) \msun$, the respective mean timescales are
$\langle t_E \rangle \simeq (50,35,3)$ days. For all curves, the
optical depths are $\sim 1.5 \times 10^{-6}$, consistent with the
theoretical calculations above and the observations.
The inclusion of the {\Nomad} population does not affect the total optical
depth because this quantity is independent of the mean mass of the
lens population.

\subsection{All-sky event rate} 

We now move on to examining the all-sky event rate distribution. In
addition to the ingredients input into Eq.~\ref{eq:microlensing_rate},
here we require two additional pieces of information: the luminosity
function of sources, $\phi(m)$, where $m$ is the source apparent
magnitude, and the radial distribution of sources $n_S(r)$.  For the
former we use the solar neighborhood $V$-band luminosity function as
compiled in \citet{Binney1998}, and the $V$-band dust extinction model
for the Galaxy as parameterized in~\citet{Belokurov:2001vh}. For the
latter, we scale the disk density profile by the local density
$\rho_0$, i.e. $n_S = \rho_d/\rho_0$.  Note that here we exclude bulge
sources because they only have a few percent
contribution to the all-sky microlensing event rate.

With the definitions above, the total microlensing event rate brighter
than a limiting magnitude, $\mlim$, is
\begin{eqnarray} 
\Gamma(< \mlim) & = & u_T
      \int_0^{m_{\rm lim}} \phi (m) \, dm 
      \int_0^\infty dD_S D_S^2 n_S(l,b,D_L)   \nonumber \\
  &\times &   
      \int_0^{D_S} dD_L \int dv_l dv_b v  f(v_b,v_l) \nonumber \\
  &\times& 
      \int_{\mcut}^{\infty} dM \zeta (M) \int R_E \rho_L(l,b,D_L) \, .
\label{eq:allsky_microlensing_rate}
\end{eqnarray} 

Figure~\ref{fig:event_rate} shows the integrated all-sky event rate as
a function of the limiting magnitude, for the same sets of {\Nomad} and
brown dwarf mass function parameters that are shown in
Figure~\ref{fig:te}. Here we have included only the event rate for
30 minutes $< t_E <$ 1 day. The lower
bound for this timescale distribution is motivated by considering the
mean timescale for an object of mass $10^{-8} \msun$, while the upper
bound is motivated from Fig.~\ref{fig:te}, which shows that events
from objects with mass $< 10^{-2} \msun$ predominantly have timescales
$\lesssim 1$ day. We will further motivate the lower cut-off of
30 minutes when we discuss analysis of observational prospects
below. For each curve, the value of $\mcut$ is indicated.  As
Figure~\ref{fig:event_rate} shows, there is $\sim 4$ orders of
magnitude uncertainty in the predicted {\Nomad} event rate brighter than
20th magnitude.  For the most shallow allowable {\Nomad} mass function,
$\alphapl = 0.5$, the event rate in this range of timescales is $\sim
0.2$ per year, while for the steepest allowable mass function,
$\alphapl = 2$, the event rate is $\gtrsim 10^{3}$ per year.

We note that, when restricting to lens masses $\gtrsim 10^{-2} \msun$,
the event rates determined in Fig.~\ref{fig:event_rate} are consistent
with prior estimates of $\sim$ few per year for $V < 15$
\citep{Nemiroff:1998md, Han:2007pt}. Including the entire population
of {\Nomads}, stars, and remnants, in fact we estimate $\sim 2500$
photometric microlensing events for sources greater than 20th
magnitude. Again the vast majority of these events are from disk
sources from the high density region towards the Galactic center, with
a few percent contribution from bulge sources. The challenge for
future observations will clearly be to achieve the appropriate
efficiency to extract these short timescale events.
 
 \section{Forecast Methodology}
\label{sec:projections}

The results from the section above provide an estimate of the {\Nomad}
event rate, independent of the survey specification. In this section,
we use the above predictions to estimate how well the {\Nomad} population
can be measured, given some basic input variables for a survey.

As a general strategy, we would like to determine 
the constraints on
$\alphapl$,
$\alphabd$ and $\mcut$ 
likely to be available from
surveys of varying cadence, exposure, and
sky coverage.  Here we define the exposure in a standard manner as the
number of stars monitored, $N_\star$, during an observational time
period, $T_{\mathrm{obs}}$. For the given exposure and the Galactic model
discussed above, we take the data as the observed timescale
distributions for a set of microlensing events. We assume that there
are $n$ bins distributed over the range of observed $t_E$. The minimum
and maximum detectable timescale for a survey is set by the detection
efficiency, which we estimate below for surveys of different cadence
and exposure.

We denote $\Gamma_\imath$ as the mean event rate in the $\imath^{th}$
$t_E$ bin for a specified exposure, where $\Gamma_\imath$ is a
function of the model parameters $\alphapl$, $\alphabd$ and $\mcut$.
Our goal is to estimate how well we can measure these parameters for
an observed event rate distribution. All of the other parameters, such
as the local stellar density, the disk scale length and scale height,
the bulge mass distribution, and stellar mass function in the regime
of main sequence stars and above are fixed to their fiducial
values. We do this primarily for simplicity in order to effectively
isolate the impact of the {\Nomad} population.  We assume that the
probability for obtaining $N_\imath$ events in the $\imath^{th}$
timescale bin follows a Poisson distribution with a mean $\mu_\imath =
T_{\mathrm{obs}} N_\star \Gamma_\imath$. For the assumptions above, we
can define the elements of the inverse covariance matrix as
\begin{equation} 
F_{ab} = \sum_{\imath=1}^n \frac{T_{\mathrm{obs}} 
         N_\star}{\Gamma_\imath}  
         \frac{\partial \Gamma_\imath}{\partial \theta_a} 
         \frac{\partial \Gamma_\imath}{\partial \theta_b}, 
\label{eq:likelihood} 
\end{equation} 
where the indices $a$ and $b$ represent the model parameters, which in
our case are $\alphapl$, $\alphabd$, and $\mcut$. From
Eq.~\ref{eq:likelihood}, the one-sigma uncertainty on parameter $a$ is
${\bf F}_{aa}^{-1}$, evaluated at the fiducial values for the
parameters. To evaluate Eq.~\ref{eq:likelihood} we choose the number
of bins $n$ to be equally spaced in log intervals. The main constraint
on the bin size will be to ensure that they are wide enough to
accommodate a 50\% uncertainty in the reconstructed~$t_E$.

\section{Detection Efficiency}
\label{sec:efficiency} 

In the analysis above, we assumed 100\% efficiency when detecting 
{\Nomads} over
the entire range of event timescales. In order to obtain a more
realistic event rate for a specific survey, we must gain an
understanding of how the detection efficiency scales as a function of
event timescale. In this section, we describe the basic set-up for our
efficiency simulations, and how they are adapted to specific surveys
in the sections that follow.

We estimate the detection efficiency via a basic procedure for
generating microlensing events. We begin by drawing source and lens
objects from the appropriate disk or bulge density distribution. The
relative transverse velocity is then drawn from the velocity
distribution~\citep{Han:1995nt}.  We draw the impact parameter for the
source and lens randomly on a uniform interval out to the 
Einstein radius, and
the peak timescale of the event, $t_0$, uniformly during the duration
of a given survey, $T_{obs}$.

The above set of parameters, $(D_S, D_L, v, t_0)$, along with the
event timescale $t_E$ fully describe the microlensing event. For this
set of parameters, we compute the amplification of the source as a
function of time, which by definition peaks at~$t_0$. The
amplification is calculated at time-steps specified by the cadence of
the survey. Motivated by the two different survey set-ups that we
discuss below, we consider two different models for the survey
cadence. First, we consider a uniform cadence model in which
the number of time-steps is simply $T_{obs}/\tcad$, where $\tcad$ is
the cadence of the survey. Second, we consider a quasi-uniform
cadence, in which there are a total of $n_e$ epochs for the survey,
and $n_m$ measurements uniformly spaced per epoch. As discussed below
this is most relevant when discussing results for the~{\GAIA} survey.

For each point on the lightcurve, the error is estimated from the
expected photometric precision. Since the focus of our analysis is on
bright microlensing events, we take the error to be uniform for all
source stars, and characteristic of the survey that is considered.  We
then simulate a lightcurve point by sampling from a normal
distribution centered on the true point with a variance given by the
photometric error. The specific error assumed for each survey will be
provided below.

With the above procedure in place, it remains to quantify a criteria
for detection for a microlensing event; we choose a
relatively simple one that is appropriate for the scope of this
work. For our primary analysis we demand that three consecutive points
on the lightcurve are $ > 3\sigma$ deviations from the mean baseline
magnitude of a star. This has been used in
previous studies~\citep{Griest:2011av}, and 
provides a good approximation to 
the criteria discussed
in~\citet{Sumi:2011kj}. The detection efficiency for an input
timescale is then the ratio of the number of simulated events
that pass the selection criteria to the total number of simulated
events at the input timescale.

 \begin{figure*}
\begin{center}
\begin{tabular}{c}
{\includegraphics[height=6.5cm]{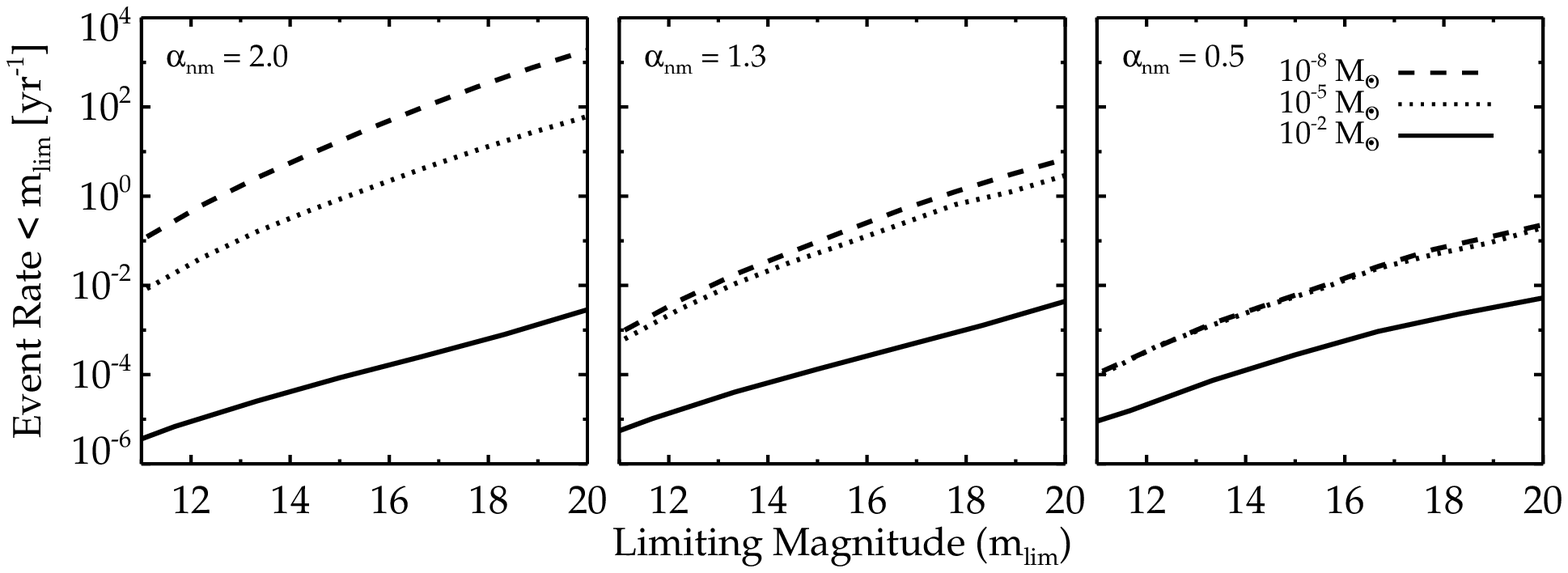}}\\
\end{tabular}
\end{center}
\caption{Event rate for sources brighter than a limiting magnitude,
  for different assumed mass functions and cadences. {\em Left} panel
  assumes the {\Nomad} and brown dwarf mass function is described by
  $\alphapl = 2$ and $\alphabd = 0$; {\it middle} panel assumes
  $\alphapl = 1.3$ and $\alphabd = 0.48$; {\it right} panel assumes
  $\alphapl = 0.5$ and $\alphabd = 1.0$. In each panel, the event rate
  is integrated between timescales of $30$ minutes $< t_E < 1$ day,
  and the value of $\mcut$ for each curve is indicated in the upper
  right panel.
\label{fig:event_rate}
}
\end{figure*}

\section{Projections and Constraints for Specific Surveys} 
\label{sec:surveys}

With the above ingredients in place, we now move on to discussing
event rates and constraints for specific surveys. We begin by
examining next-generation bulge surveys with {\WFIRST}, and then move
on to discuss forthcoming all sky surveys {\GAIA} and {\LSST}. We
conclude by examining the detection prospects 
in the short term for the Kepler satellite.

\subsection{WFIRST}

We first consider the case of a dedicated survey to monitor the inner
Galaxy region. This is similar in spirit to the modern MOA, OGLE, and
EROS surveys, and to a larger scale, space-based extension such as the
proposed {\WFIRST} mission~\citep{Green:2011zi}.  For the
former set of surveys, we can directly use their published detection
efficiencies to predict the event rates and model the error
distributions, while for a~{\WFIRST}~type mission this requires
simulating events as described above.

For {\WFIRST}, we use a cadence of 15 minutes, a total exposure time
of 1 year, and photometric errors of $0.1\%$, which will be achievable
down to $J = 20.5$. Using the above model, at $t_E \sim 0.03$ days we
find a detection efficiency of $\sim 50\%$. This high efficiency at
short timescale is primarily driven by the order of magnitude increase
in the photometric precision relative to modern microlensing
experiments.  We note that if we assume the MOA-II cadence and
photometric uncertainty in their high cadence fields, which we
approximate as $\sim 30$ observations per night at $\sim 15$ minute
cadence, at $t_E = (0.5,1,10)$ days, we find efficiencies of
$(10,20,40)\%$, which provides a good approximation to the efficiencies
reported in \citet{Sumi:2011kj}.

\begin{figure*}
\begin{center}
\begin{tabular}{cc}
{\includegraphics[height=7.cm]{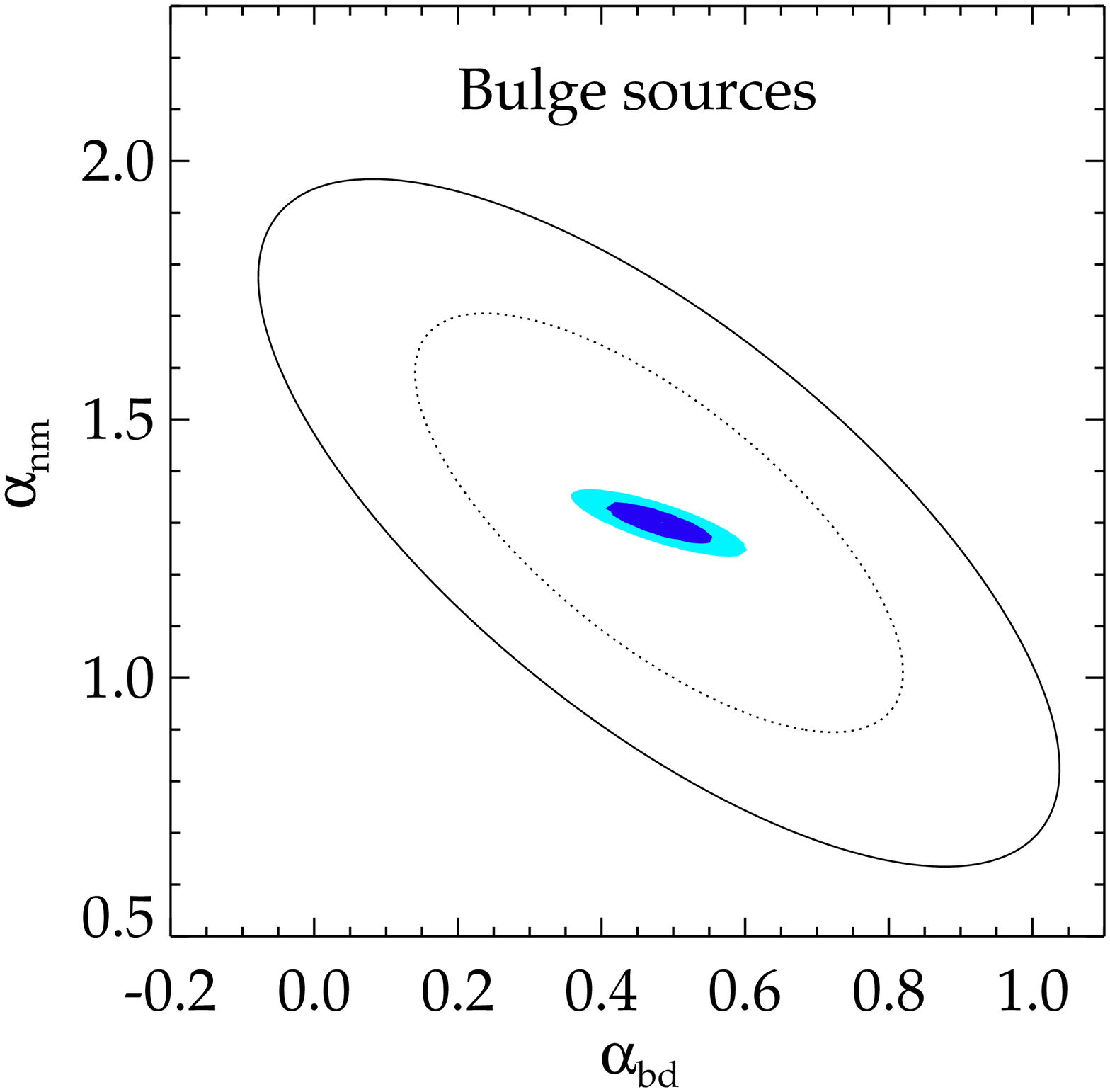}} & 
{\includegraphics[height=7.cm]{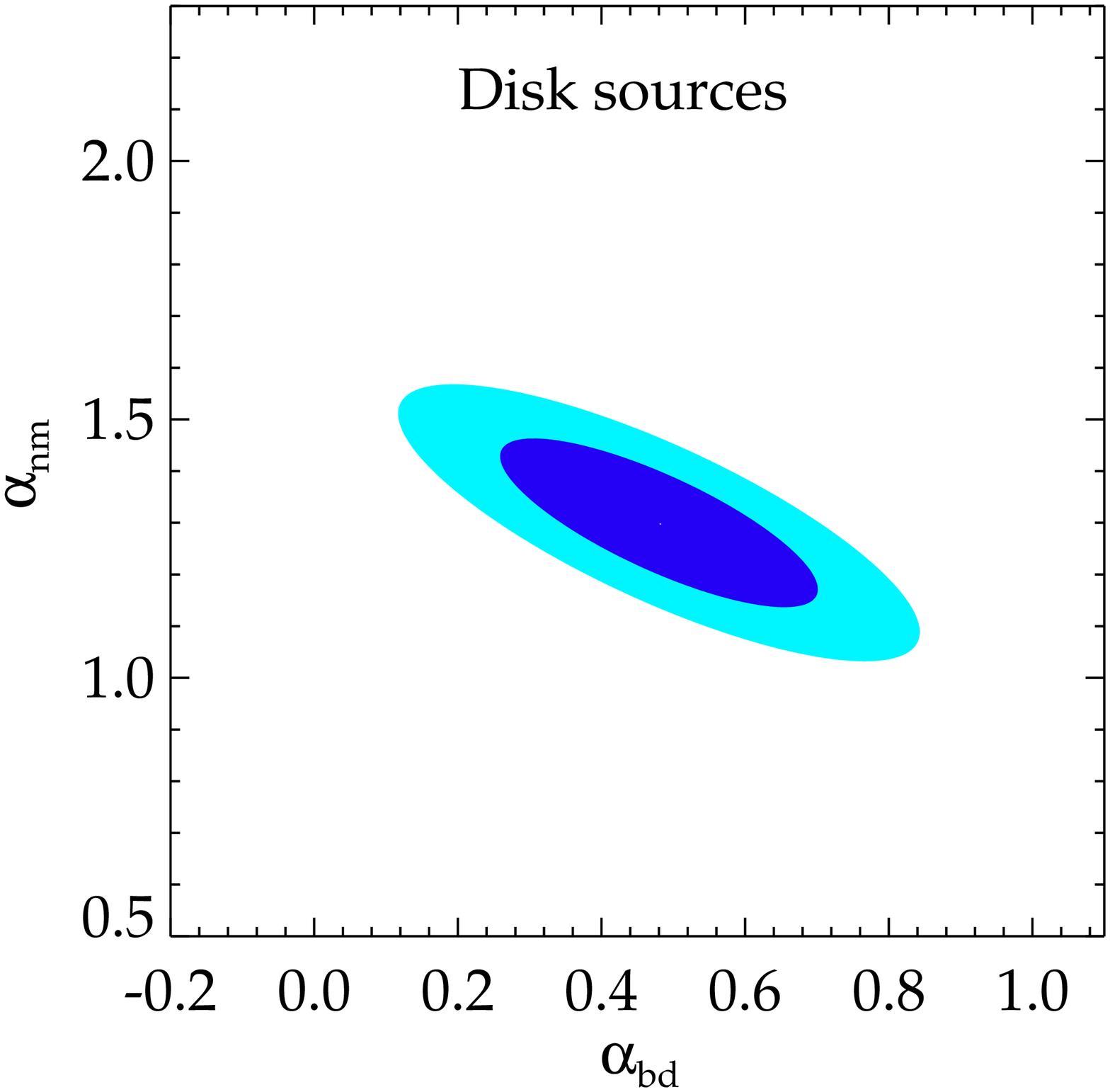}} \\
\end{tabular}
\end{center}
\caption{{\em Left Panel}: Joint constraints on the slope of the mass
  function in the {\Nomad} region, $\alphapl$, and the slope in the low
  mass stellar regime, $\alphabd$ for bulge sources. 
  Unfilled contours
  assume the exposure and efficiency of modern bulge surveys, and
  match the results from~\citet{Sumi:2011kj}. Inner contour is 68\%
  c.l. and outer contour is 95\% c.l. Filled contours are the
  projected constraints for an exposure and efficiency expected for
  {\WFIRST}. {\em Right panel}: Similar to filled contours on the
  left, except assuming disk sources.
\label{fig:wfirst}
}
\end{figure*}

In Figure~\ref{fig:wfirst} we show the resulting one- and two-sigma
uncertainties on the combination $\alphapl$-$\alphabd$, for modern and
for future dedicated surveys. Here we have assumed fiducial values of
$\alphabd = 0.48$ and $\alphapl = 1.3$, though we generally find that
our results are insensitive to the specific value for these
quantities. In plotting the unfilled contours in the left panel, we have assumed an
exposure and detection efficiency similar to the MOA-II analysis,
which provides a total of $\sim 500$ events for 2 years of
observations of 50 million stars; we have assumed $n = 20$ bins
distributed uniformly in log between timescales of 1-200 days. In this
case the errors from our model are in good agreements with the
one-sigma uncertainties on $\alphabd$ and $\alphapl$ presented
in~\citet{Sumi:2011kj}, with slight departures due to the non-gaussian
behavior in the tails of the results from the later. In this case the
one-sigma errors are $\sigma_{\alphabd} \simeq 0.30$ and
$\sigma_{\alphapl} \simeq 0.40$.

The filled set of contours in the left panel of Figure~\ref{fig:wfirst}
show the projected constraints assuming a cadence of 15 minutes and $2
\times 10^8$ monitored stars for one year. This cadence and exposure
is motivated by the preliminary specifications for
{\WFIRST}~\citep{Bennett:2010xm}.  To provide the most optimistic
predictions, and as motivated by the photometric precision and the
simulations described above, here we have assumed a 100\% detection
efficiency at all $t_E > 0.04$ days. In this case, the one-sigma
uncertainties are reduced to $\sigma_{\alphapl} = 0.03$ and
$\sigma_{\alphabd} \simeq 0.05$, representing nearly an order of
magnitude improvement relative to the modern constraints. If we 
assume $\mcut = 10^{-3} \msun$, this corresponds to 
a measurement of $\beta$ to $\sim 13\%$ precision, and for 
$\mcut = 3 \times 10^{-7} \msun$ we have a measurement of 
$\beta$ to $\sim 25\%$ precision. 

For comparison to the bulge results, 
in the right panel of Figure~\ref{fig:wfirst} we show the resulting
constraints for disk observations towards $(-2.4^\circ, 331^\circ)$.
This direction is specifically chosen to compare to the results of 
\citet{Rahal:2009yt}. In this case, the constraints on the combination 
$\alphapl$-$\alphabd$ are $\sim 3$ times weaker primarily because 
in this direction only disk lenses are contributing to the event rate.

How well can we determine the minimum mass of a {\Nomad}, $\mcut$, from a
{\WFIRST} type survey?  Because a given mass {\Nomad} will
produce events over a fixed range of timescales (for an assumed
velocity distribution function) the answer to this question depends on
the value of $\mcut$ itself. If the mean event timescale at a given
$\mcut$ is significantly less than the cadence of the survey, then 
observations will not effectively be able to probe this mass scale.

\begin{figure}
\begin{center}
\begin{tabular}{c}
{\includegraphics[height=7.cm]{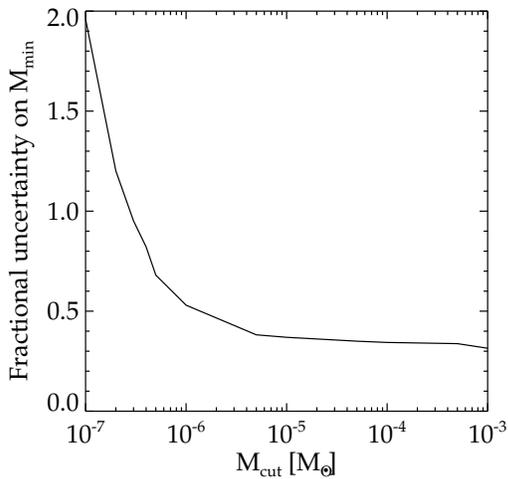}} \\
\end{tabular}
\end{center}
\caption{Fractional uncertainty on $\mcut$ for a cadence of 15 minutes and bulge
  observations.
\label{fig:mcut} 
}
\end{figure}

In Figure~\ref{fig:mcut} we show the resulting fractional uncertainty
on $\mcut$ for a cadence of 15 minutes and bulge observations.  In
this case, for $\mcut \gtrsim 10^{-5} \msun$ we find fractional
uncertainty $\gtrsim 30\%$. In fact down below the Earth mass scale
for $\mcut \gtrsim 10^{-6} \msun$ we still find fractional uncertainty
$\sim 50\%$, below which there is degradation of the constraints
because the event rate in the observable timescale window becomes too
low.

\subsection{Large-Scale Surveys: {\GAIA} and {\LSST}} 

We now extend to consider projected constraints on
$\alphapl$-$\alphabd$ from all-sky observations. In this case,
estimating the detection efficiency of the survey will be crucial in
order to understand what fraction of the total event rate shown in
Figure~\ref{fig:event_rate} will be accessible.

The two primary templates we consider for large scale surveys are
those being planned for
{\GAIA} and {\LSST}. 
These multi-purpose surveys are not
expected to have cadence as high as the dedicated inner Galaxy
observations discussed above, so they will not be as sensitive to very
short timescale microlensing events. However by their nature all-sky
observations do probe the {\Nomad} population on a Galaxy-scale that
are inaccessible to dedicated pointings towards a fixed region of the
Galaxy.

\subsubsection{{\GAIA}} 
\label{ssec:GAIA}

As our first example of a survey with a non-uniform cadence, we
consider {\GAIA}, which is scheduled to launch in 2013. Though {\GAIA} 
is primarily designed as an astrometric mission, it will have a single 
measurement photometric accuracy of $\sim 10$~mmag for sources brighter
than its broadband 20th magnitude.
Because of the {\GAIA} scanning strategy, the sampling for each star
is not uniform during the mission lifetime. Measurements will be
grouped into epochs, during which an observation is performed in $\sim
6$ hr intervals. The mean number of measurements per epoch is $\sim
5$, though some epochs will have a minimum of $7$ measurements
~\citep{Eyer:2005fa}. The mean number of visits between epochs is
25-35 days, though depending on Galactic latitude we estimate from the
results of \citet{Eyer:2005fa} that $\sim 10\%$ of the stars will have
$\sim 5$ days between epochs.

Motivated by these specifications, we model {\GAIA} observations by 
considering a quasi-irregular sampling pattern.  For the baseline model 
we assume 25 days between epochs, and within each epoch there are 
five photometric measurements. This is the approximate mean sampling 
rate of {\GAIA}~\citep{Eyer:2005fa}. The survey is run for a total of $T_{obs} =
5$ years, resulting in a mean of 300-400 photometric samples for the
lifetime of the survey. To model the distribution of disk sources we use the 
$V$-band luminosity function described above, along with the
\citet{Belokurov:2001vh} dust extinction parameterization.

For a {\GAIA}-like sampling, 
the short timescale events, $t_E \sim 1$ day, will occur {\it during}
an epoch, and it is possible that a peak of the microlensing event 
will not be
discernible. To account for this, we modify the detection criteria. 
For a simulated event at an input timescale, we again search for three 
points on the lightcurve that are greater than 3-$\sigma$ deviations from 
the baseline magnitude of the source.
In addition we include a second, stricter cut to the detection criteria,
namely that the peak of the event is observable. 

Given the above algorithm, for the~{\GAIA}~sampling pattern, at $t_E =
1$ day we find a detection efficiency of $1\%$.

For the~{\GAIA}~cadence and estimated efficiency, in
Figure~\ref{fig:allsky_contour} we show the joint constraints on
$\alphapl$ and $\alphabd$ for all-sky observations. Here we have
assumed a five year lifetime of the mission.  In this case the
constraints are similar to the current 
constraints on these parameters
because of the similar event rates after our detection efficiency cuts
have been implemented.

\subsubsection{{\LSST}} 
\label{ssec:LSST}

As our second example, 
we examine the somewhat deeper survey we anticipate being carried out by 
the {\LSST}. This system will repeatedly survey the entire
visible southern sky to a $5$-sigma point source depth of 
$g=25.0$, $r=24.7$
per epoch. {\LSST} is expected to have a mean cadence (across all filters) 
of less than~4 days
and a mission lifetime of 10 years \citep{:2009pq}.
To achieve a synoptic survey, the {\LSST} will follow a logarithmic sampling
pattern, with 15-second exposures separated by 15 seconds, 30 minutes, 3--4 days
and one year, with considerable scatter in the two intermediate cadences to
allow flexible scheduling. Two back-to-back exposures constitute a ``visit'';
the baseline plan has each field being visited twice on any of its
observation nights. As with {\GAIA}, detection of a {\Nomad} by microlensing
requires seeing both sides of a peak in the lightcurve, suggesting that events
with timescales less than 3 days may be difficult to detect.

To approximate the sampling strategy of {\LSST}, we assume a uniform
cadence for the lifetime of the survey.
From the formalism above we 
have calculated the detection efficiency for cadences of both 1 and 4 
days; we consider higher cadence dedicated campaigns with {\LSST} below.
We find that only for a 1 day cadence is it possible to achieve 
1\% detection efficiency for timescales $t_E \gtrsim 1$ day. For a 
4-day cadence, the efficiency for detecting {\Nomads} drastically drops
(though in this case a large number of brown dwarf events will still be 
measured very precisely). We utilize this
optimistic 1\% efficiency when we calculate the projected constraints on
$\alphapl$-$\alphabd$ for the uniform cadence model. 

To model the distribution of sources for our {\LSST} predictions we
use the $I$-band luminosity function from \citet{Zheng:2003nc}. In
this case, dust extinction modeled by adopted in the model
of~\citet{Belokurov:2001vh} and scaling according to the standard
extinction law between wavebands \citep{Rieke1985}.

\begin{figure*}
\begin{center}
\begin{tabular}{cc}
{\includegraphics[height=7.cm]{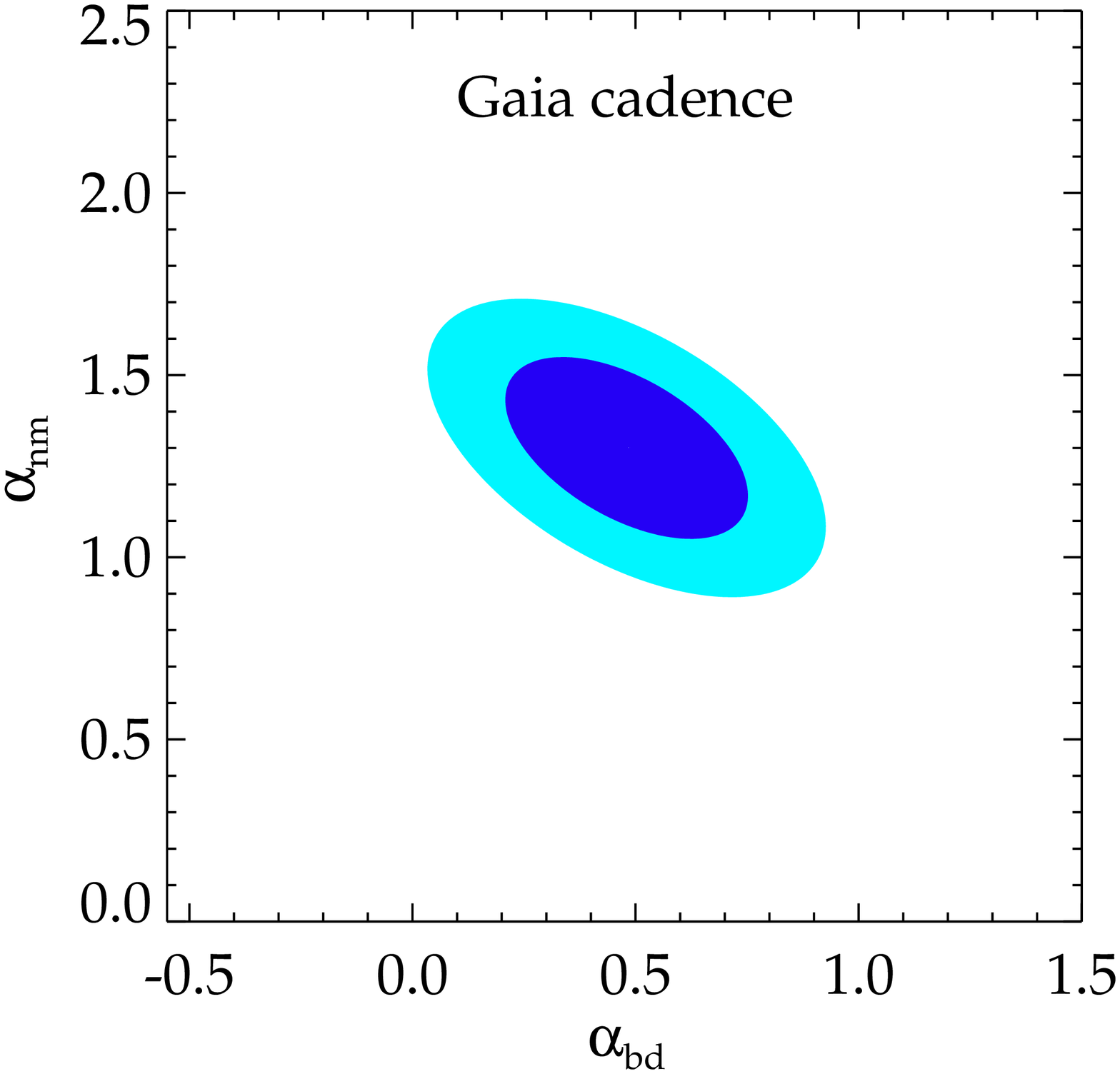}} & 
{\includegraphics[height=7.cm]{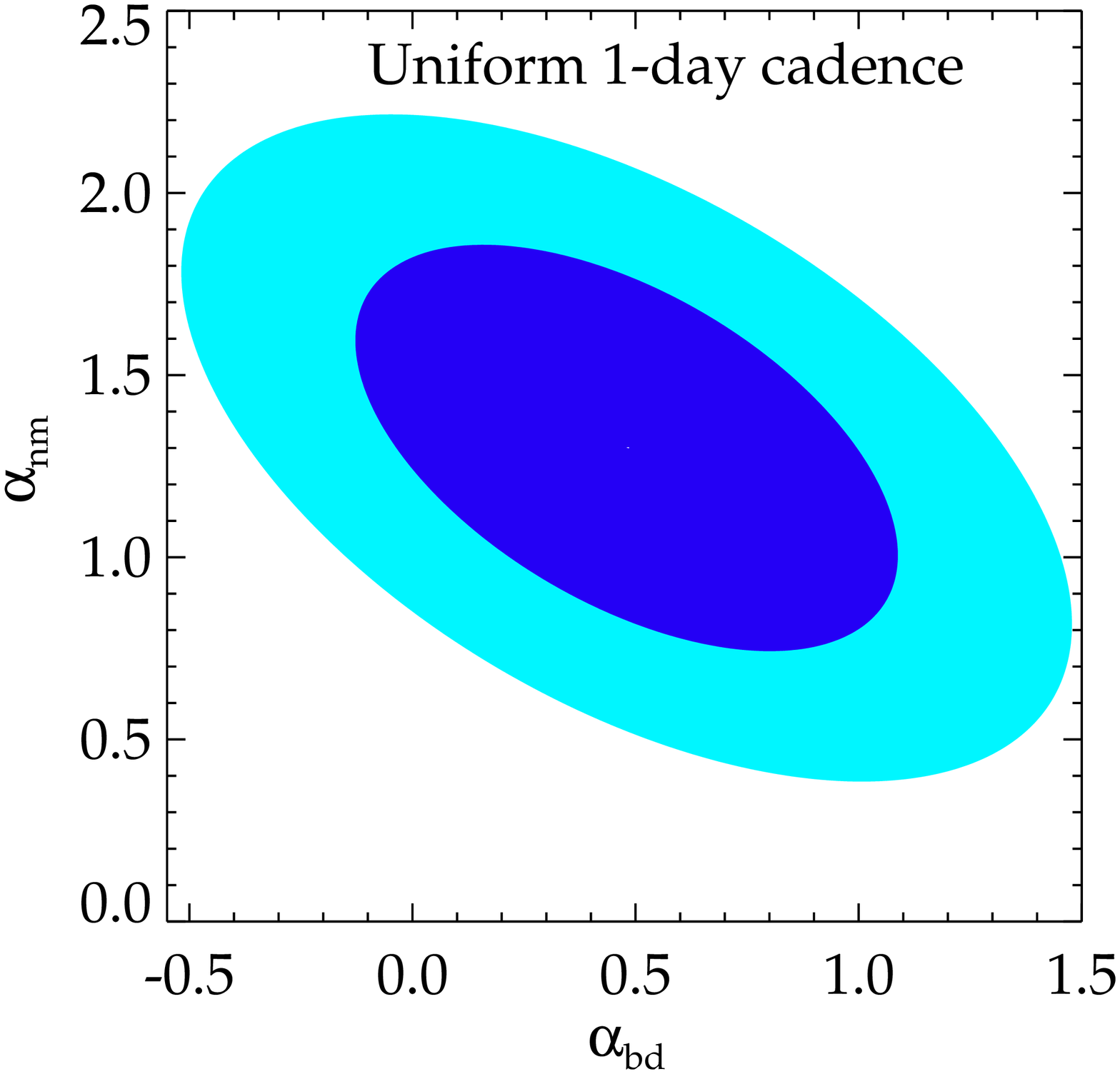}} \\
\end{tabular}
\end{center}
\caption{Joint constraints on $\alphapl$ and $\alphabd$ for all-sky
  observations.  The left panel shows results for the irregular
  sampling pattern that represents the~{\GAIA}~mission, while the
  right panel represents the uniform sampling pattern that may be
  achievable with~{\LSST}.
\label{fig:allsky_contour}
}
\end{figure*}

The left panel of Figure~\ref{fig:allsky_contour} shows the results of the analysis. As
indicated, the constraints are weaker relative to the left panel; this is mainly
due to the reduced efficiency as compared to the {\GAIA} sampling model.  

\subsection{Kepler}
\label{ssec:kepler}

As our final example we study the {\Nomad} event rate measurable by the 
Kepler satellite.\footnote{http://kepler.nasa.gov/} 
Kepler monitors $\sim 100$ deg$^2$ towards the 
Cygnus region, and has a photometric precision of approximately 80 ppm for sources 
brighter than $V=13$, and a few percent for sources at $V=20$. Kepler is
complete down to $V=17$. The integration time is 30 minutes for the
majority of Kepler sources.  
For $\alphapl = 2$ and $\mcut =
10^{-8} \msun$, and assuming $\mcut = 10^{-8}$ down to 20th magnitude,
the raw event rate is a few per year for 30 minutes $< t_E <$ 1 day.
For $\alphapl < 2$ we find less than one event per year. However,
these predictions are for $u_T = 1$; due to its precise photometry of
$\sim 80$ ppm for bright stars with $V < 13$ the event rate is
proportionally larger for $u_T > 1$. Thus discovery of anomalous
microlensing events in the Kepler data may indicate a steep value for
the {\Nomad} mass function, and warrant a dedicated analysis of the
photometry of Kepler stars.

\section{Detecting Short Timescale Events}
\label{sec:drift} 
In the above analysis we have restricted only to
events with timescales sufficiently long to be detectable according to
the criteria above. What if we relax this criteria, and expand to
consider events with shorter timescales, over which the lightcurves
are much more sparsely sampled? Is it still possible to detect 
microlensing events from lighter nomads over much shorter timescales?

As an example let us consider the planned survey strategy
of~{\LSST}. In a given~{\LSST} filter, each visit will consist of two
consecutive 15~second exposures separated by a 4~second readout
interval.  When possible, each field will be observed twice, with
visits separated by $\sim 15-60$ minutes.  For stars with $r \lesssim
20$, the single visit photometric precision of each measurement is
$\sim 10$ mmag. Though per visit the photometry is very precise and
the 30 second cadence is short, two points on a lightcurve are not
adequate to claim the detection of a microlensing event.  However, the
lack of variation of a source star over this timescale in between
visits could allow us to bound the existence of {\Nomads} with
characteristic timescales $\sim 30$ seconds.  Assuming both the lenses
and the sources to be in the disk, this timescale corresponds to a
lens of mass $\lesssim 10^{-10} \msun$.

Will lenses with such a small mass cause noticeable brightness
fluctuations in a star when accounting for finite source effects? For
our uniform source surface brightness model, we find that this depends
on the lensing geometry. For example, with $D_L = 1$ kpc and $D_S = 5$
kpc, a lens of mass $[10^{-9}, 10^{-10}] \msun$ will have a peak
brightness of $\Afs \simeq [1.01, 1.001]$. For lenses nearer to the
source, $\Afs$ is reduced from these values. Though these are smaller
brightness fluctuations than typically searched for in microlensing
events, the presence of these objects may be limited 
given the photometric precision of {\LSST}. 

In perhaps more near of a term, we may entertain the prospect of a 
dedicated telescope that is capable of detecting short timescale microlensing 
events that last for as little as tens of seconds. As an example, we will consider a liquid
mirror telescope similar in design to the six-meter Large Zenith Telescope
~\footnote{http://www.astro.ubc.ca/lmt/lzt/}, 
only in our case positioned in the Southern hemisphere to cover the 
Galactic bulge. For a $4000 \times 4000$ CCD chip with $1^{''}$ per pixel and a 
6-m aperture, in less than a day a patch of area $\sim 100$ square degrees could be scanned. 
We focus on the $I$-band, and take the bulge as an example with an surface
brightness of $17.6$ mag arcsec$^{-2}$~\citep{1988AJ.....96..884T}. Assuming that the signal-to-noise is
dominated by the unresolved light from the bulge and shot noise, for a star with $I=19$
the signal-to-noise is $S/N \simeq 37\sqrt{t/{\rm sec}}$. A 10-second 
exposure then 
gives a photometric precision of $\sim 1\%$, and during this time a star crosses
through $\sim 10^3$ pixels. This would likely be sufficient to obtain several points on 
a lightcurve to measure a microlensing event with $t_E \sim 30$. sec. 

For a nomad mass function of $dN/dM \sim M^{-2}$ with $\mcut = 10^{-9} \msun$, 
for a 100 deg$^2$ patch that passes through the Galactic center we find $\sim 50$ events per year 
with $t_E > 30$ seconds for source stars with $I < 19$. The event rate may even be up to an 
order of magnitude larger for steeper values of the mass function over the range $10^{-9} \msun$. 
It is also worthwhile to point out that a telescope designed along these lines could also be a relatively 
inexpensive endeavor. Further, a liquid mirror telescope with mercury could extend to the near-infrared,
where the reflectivity would be similar to that in the optical. 

\section{Discussion and Conclusions}
\label{sec:conclusion}

We have estimated that there may be up to about $10^5$ compact objects 
per main sequence star in the Galaxy that are greater than the mass of Pluto. 
A dedicated high cadence survey of the inner Galaxy, 
such as would be possible with {\WFIRST}, could 
measure the number of nomads greater than the mass of Jupiter
per main sequence star to $\sim 13\%$ , and the corresponding number 
greater than the mass of Mars to $\sim 25\%$. 
Also {\WFIRST} can measure the minimum mass of 
the {\Nomad} population to about 30\%. Large-scale surveys, 
in particular that of {\GAIA}, could
identify {\Nomads} in the Galactic disk that are greater than about the mass of Jupiter. 

Observations along the lines that we discuss will constrain the {\Nomad} population of the disk
relative to the bulge, and will also more generally improve the
star-star microlensing event rate in the disk and the solar
neighborhood, about which very little is now
known~\citep{Gaudi:2007xa,Fukui:2007gi,Rahal:2009yt}.  Further, they
will improve our understanding of the mass function of low mass
brown dwarfs and super-Jupiters, and the
distinction between these classes of objects~\citep{Spiegel:2010ju}.

How will these measurements compare to modern microlensing measurements of
low mass brown dwarf population from disk observations? To answer this
question we can briefly consider the results from
\citet{Rahal:2009yt}.  
These authors find a total of $\sim 20$ events in
three fields in which the lenses are primarily disk sources, and in
particular there are two very short timescale lenses, at $t_E = 7$ and
$12$ days.  While the data are not sufficient at present to perform a
full statistical analysis and constrain $\alphapl$ and $\alphabd$,
from an analysis of these data one may deduce that a steeper model
brown dwarf mass function is favored over a more shallow model.  The
inclusion of the {\Nomad} population does mildly improve the statistical
fit, though in order to probe this population with disk observations a
survey must build up a sufficient event rate in the $\sim 1-10$ day
timescale bin.

How will the future microlensing measurements we discuss compare to direct measurements 
of the brown dwarf mass function? \citet{Metchev:2007sy} find that for warm brown
dwarfs the mass function may be flat, $\alphabd = 0$. For cooler brown
dwarfs the recent WISE observations are consistent with a wide range
of $\alphabd$ between $0-1$ \citep{Kirkpatrick:2011tg}.  Other
microlensing observations shed light on the brown dwarf mass function,
though they do not clarify the picture. For example,
\cite{Gould:2009ku} uncover an extreme magnification microlensing
event and interpret it as due to a thick disk brown dwarf. According
to ~\cite{Gould:2009ku}, there is a very low probability to observe
this event given our standard population of brown dwarfs in the
Galactic disk given the large velocity of the event.  The existence of
these events either implies that we have been lucky to observe events
at all (in particular with the large observed magnifications), or that
the local population of low mass and low luminosity stellar remnants
is larger than is presently predicted. 

If a nomad is identified via the methods
described in this paper, there are a number of follow-up observations
that are possible. For example even though {\GAIA} will only do on
average 1-2 one-dimensional astrometric measurements per epoch, it may
be possible to confirm the photometric detection with astrometry for
the brightest sources by comparing the centroid of the source during
the event to the baseline centroid as determined over several epochs
during the course of the mission.  

Finally we note that an additional outcome of the 
observational approach discussed above, especially regarding the 
detection of short timescale microlensing events, is that upper limits may be 
set on the density of {\Nomads}. This could set very interesting 
constraints on the population of planetesimals in nascent planetary 
systems. 

\section*{Acknowledgements}
We acknowledge Ted Baltz for several useful discussions during the course 
of this work. LES acknowledges support for this work from NASA through Hubble
Fellowship grant HF-01225.01 awarded by the Space Telescope Science
Institute, which is operated by the Association of Universities for
Research in Astronomy, Inc., for NASA, under contract NAS 5-26555. LES
acknowledges additional support by the National Science Foundation
under Grant No. 1066293 and the hospitality of the Aspen Center of
Physics. M.B. acknowledges support from the Department of Energy
contract DE-AC02-76SF00515. PJM acknowledges support from the Royal Society in
the form of a University Research Fellowship.


\end{document}